\documentclass[aps,prb,twocolumn,superscriptaddress,amsmath,amssymb,showpacs]{revtex4}
\usepackage{graphicx}% Include figure files
\usepackage{dcolumn}% Align table columns on decimal point
\usepackage{bm}% bold math

\newcommand{\be}{\begin{equation}} %equation commands
\newcommand{\ee}{\end{equation}}
\newcommand{\ba}{\begin{eqnarray}}
\newcommand{\ban}{\begin{eqnarray*}}
\newcommand{\ean}{\end{eqnarray*}}
\newcommand{\ea}{\end{eqnarray}}
\newcommand{\ket}[1]{\left|\, #1\, \right\rangle} %bra-ket notation

\newcommand{\bra}[1]{\left\langle\, #1\, \right|}

\begin{document}
\title{Quantum Noise, Scaling and Domain Formation in a Spinor BEC}
\author{George I. Mias}\email{george.mias@aya.yale.edu}
\affiliation{Sloane Physics
Laboratory,Yale University, New Haven, CT 06520-8120, USA}
\author{Nigel R. Cooper}
\email{nrc25@cam.ac.uk} \affiliation{T.C.M. Group, Cavendish Laboratory, J. J. Thomson Ave., Cambridge, CB3 0HE, UK}
\author{S. M. Girvin}
\email{steven.girvin@yale.edu} \affiliation{Sloane Physics
Laboratory,Yale University, New Haven, CT 06520-8120, USA}

\date{\today}
\begin{abstract}  In this paper we discuss Bose-Einstein spinor condensates for F=1 atoms in the context of $^{87}$Rb, as studied experimentally by the Stamper-Kurn\cite{Sadler:2006} group. The dynamical quantum fluctuations of a sample that starts as a condensate of $N$ atoms in a pure $F=1$, $m_F$ = 0 state are described in analogy to the `two-mode squeezing' of quantum optics in terms of an  $\mathfrak{su}$(1,1) algebra.  In this system the initial $m_F=0$ condensate acts as a source (`pump') for the creation pairs of $m_F$ =1,-1 atoms.  We show that even though the system as a whole is described by a pure state with zero entropy,  the reduced density matrix for the $m_F$ = +1 degree of freedom, obtained by tracing out the $m_F$ = -1,0 degrees of freedom, corresponds to a thermal state. Furthermore, these quantum fluctuations of the initial dynamics of the system provide the seeds for the formation of domains of ferromagnetically aligned spins.
\end{abstract}
\maketitle
\section{\label{Introduction} Introduction}
Spinor condensates were first
realized\cite{Stamper-Kurn:1998,Stenger:1998} in 1998. Such
condensates are very rich in the underlying physics and have been
the subject of numerous studies, from mean field theory and
ground state considerations\cite{Ho:1998,Koashi:2000,Ohmi:1998}
to experimental observation of dynamics, rotation effects,
domain structures \cite{Chang:2005a,Chang:2004,
Chang:2005,Miesner:1999,Stamper-Kurn:1999, Lamacraft:2007},
theoretical  dynamics of mixing  \cite{Law:1998,
Pu:1999,Yi:2003};  existence of quantum vortices and
topological structure \cite{Isoshima:2001, Saito:2006,
Saito:2007,Saito:2005};  dipolar effects \cite{Kawaguchi:2007}
and very many more, currently making this one of the most
active fields at the interface between condensed matter and
atomic physics.

In this paper we investigate the formation of domains in terms
of a Bose-Einstein spinor condensate of alkali atoms, $^{87}$
Rb, in the hyperfine multiplet $F=1$, which has been shown to
have ferromagnetic interaction amongst the constituent
atoms\cite{Ho:1998,Ohmi:1998,Chang:2005}.  A recent experiment
carried out by the Stamper-Kurn group at
Berkeley\cite{Sadler:2006}, showed the formation of domains in
a $^{87}$Rb gas starting with a polar non-ferromagnetic initial
state. The experimental sample formed random domains of varying
transverse magnetization, and this paper is based on
considerations regarding this experiment and the dynamical
seeding of the observed domains. The analysis of this
experiment has been undertaken by
Lamacraft\cite{Lamacraft:2007} who derived a form for the
dynamics of the particles in different states. The dynamics of the associated quantum phase transition and scaling properties of the system magnetization has been discussed by Damski et al.\cite{Damski:2007} for a one-dimensional system treated in a mean-field approach, with extensions to higher dimensions.  The recent work of Saito et al.\cite{Saito:2007a} further investigated the dynamic formation of these domains in terms of quantum noise, including a simulation of two-dimensional domains using a
combination of quantum dynamics and Gross-Pitaevskii evolution
- this work was completed concurrently with our own
investigation with essentially the same
conclusions\cite{Mias:2007a}. In this paper we discuss an
alternate, yet equivalent description of the experimental work,
which elucidates the nature of the quantum statistics. After
introducing the system and the relevant experiment in
Secs.~\ref{Introduction}-\ref{exptBK}, we proceed in
Sec.~\ref{singleMode} to supplement previous results on the
single-mode problem by providing a  connection to the ideas of
quantum squeezing and quantum noise, as well as some exact
solutions to the dynamical single-mode problem.   In
Sec.~\ref{multimode} we also follow a multi-mode approach, in
terms of which the time evolution, quantum noise and statistics of
the domain seeding may be discussed.  This is presented in
connection to an $\mathfrak{su}$(1,1) algebra inherent in the
effective Hamiltonian derived for the system,  making it
possible to obtain probability distributions for the
fluctuations in the system.  As we will see the thermal nature
of these distributions is distinctive and might be verifiable
experimentally.

\subsection{\label{SpinorTheory}Effective theory of spinor condensates }
The use of dipolar optical traps allows for Bose-Einstein
condensation of alkali atoms in which the spin degree of
freedom is still active - the traps  do not preferentially
select one of the spin states with a given $m_F$ (as opposed to
magnetic traps that favor the weak-field seeking states).  Thus
for a gas of spin $F$ there are $2F+1$ degrees of freedom
deriving from the hyperfine spin. The total spin for both
$^{87}$Rb and $^{23}$Na is the sum of the nuclear spin,
$I=3/2$, and the electronic spin $S=1/2$, leading to a total
spin $F=3/2\pm 1/2$.   For the $F=1$ manifold, which is usually
probed in optical trap experiments, this means that the atoms
may be described by a spinor,
  \begin{equation}\label{eq:3comporder}
 \text{\boldmath $\psi$}({\bf r}) = \left( \begin{array}{c}
  \psi_{1}({\bf r}) \\
  \psi_{0}({\bf r}) \\
  \psi_{-1}({\bf r}) \end{array} \right),
\end{equation}
with each component, $\psi_i({\bf r})$, corresponding to the
wave function of a species in the $m_F=\{-1,0,1\}$ state.  At
zero magnetic field the system's spin rotational invariance is
manifest. The standard Hamiltonian for the low energy dynamics
of an $F=1$ dilute atomic gas was developed by Ho\cite{Ho:1998}
and Ohmi and Machida \cite{Ohmi:1998}. This assumes that only
two-body collisions are important and that the atoms do not
interact otherwise with each other and that the system is
rotationally invariant, which means that the interactions can
only depend on the total spin $\mathcal{F}$ of the colliding
atoms and not on
$\mathcal{F}_z$\cite{Dalfovo:1999,Esry:1999,Leggett:2001,Pethick:2002}.
The complete effective Hamiltonian for the $F=1$ dilute cold
gases in second quantized form is written as, \ba
H&=& \int d^3 {\bf r}\bigg{\{} \psi^\dagger_i({\bf r})\left(-\frac{\hbar^2\nabla^2}{2m}\right)\psi_i({\bf r})+\nonumber\\
&\ &+\psi^\dagger_i({\bf r}) V_{ij}\psi_j({\bf r})+ \frac{1}{2}\bigg{[}c_0\psi_i^\dagger({\bf r})
\psi_j^\dagger({\bf r})\psi_i({\bf r})\psi_j({\bf r})\nonumber\\
&\ &+ c_2\left(\psi_i^\dagger({\bf r}){\bf F}_{ij}\psi_j({\bf
r})\right)\cdot \left(\psi_k^\dagger({\bf r}){\bf
F}_{kl}\psi_l({\bf r})\right)\bigg{]}\bigg{\}},\label{Hfull}
%:eqn Hfull
\ea where $V_{ij}$ is the trapping potential, the indices refer
to the hyperfine species, $i,j \in\{-1,0,1\}$, and we are using
the Einstein summation convention of summing over repeated
indices.  ${\bf F}\equiv\{F_x,F_y,F_z\}$ is a vector of
spin-one matrices. We have also identified the couplings \ba
c_0&\equiv&\frac{4\pi\hbar^2}{3m} (2a_2+a_0),\\
c_2&\equiv&\frac{4\pi\hbar^2}{3m} (a_2-a_0), \ea which depend
on only two parameters, the scattering lengths $a_{0}$ and
$a_{2}$. The field operators obey the commutation relations,
\be [\psi_i({\bf r}),\psi_j^\dagger({\bf
r}')]=\delta_{i,j}\delta^3({\bf r}-{\bf r}'), \ee with all
other commutators being zero. The trapping potential, $V_{ij}$
is the result of a combination of magnetic and optical
potentials and may be taken as diagonal, $V({\bf r})$, assuming
that all magnetic fields are in the `z-direction'.  We should
keep in mind that in writing the Hamiltonian in terms of
projecting onto total spin $\mathcal{F}$ subspaces we are
assuming that any Zeeman shifts are smaller than the hyperfine
splitting so that $\mathcal{F}$ is still a good quantum number.
The mean field analysis of this Hamiltonian\cite{Ho:1998,
Ohmi:1998,Yi:2003,Law:1998} indicates that the ground state for
$c_2<0$, as is the case for $^{87}$Rb, is ferromagnetic,
 \ba
\zeta_0&=&\exp[i\theta]U(\beta_1,\beta_2,\beta_3)\left(
\begin{array}{c}
1 \\
0 \\
0 \end{array} \right)\nonumber\\
&=&\exp\left[i\theta-\beta_3\right]\,\left( \begin{array}{c}
\exp\left[-i\beta_1\right]\cos\left[\frac{\beta_2}{2}\right]^2 \\
\sqrt{2}\cos\left[\frac{\beta_2}{2}\right]\sin\left[\frac{\beta}{2}\right] \\
\exp\left[i\beta_1\right]\sin\left[\frac{\beta_2}{2}\right]^2
\end{array} \right), \ea which is written this way to emphasize
that we have a degenerate set of ground states which are
related to each other by a gauge transformation,
$\exp[i\theta]$, and an arbitrary rotation via
\be
U(\beta_1,\beta_2,\beta_3)\equiv
\exp[-iF_z\beta_1]\exp[-iF_y\beta_2]\exp[-iF_z\beta_3],
\ee
where $\{\beta_1,\beta_2,\beta_3\}$ are the Euler
angles\cite{Lifshitz:1982a}.  The application of a magnetic
field would lead to the average spin pointing in the direction
of the applied field and the selection of one of the degenerate
ground states.  In the absence of a field we have spontaneous
symmetry breaking, where the system selects at random one of
the possible directions for its spins.

We may write the field operators for the case of a homogeneous
system as a plane wave expansion:
\be \psi_i({\bf
r})=\frac{1}{\sqrt{V}} \sum_{\bf k} a_{i{\bf k}}\exp[i {\bf
k}\cdot{\bf r}].
\ee
This gives:
\begin{widetext}
\ba
H_I&=&\frac{1}{2V}\sum_{{\bf k}_1+{\bf k_2}={\bf k_3}+{\bf k}_4}\bigg{\{} (c_0+c_2)
(a^\dagger_{1{\bf k}_1}a^\dagger_{1{\bf k}_2}a_{1{\bf k}_3}a_{1{\bf k}_4}
+a^\dagger_{-1{\bf k}_1}a^\dagger_{-1{\bf k}_2}a_{-1{\bf k}_3}a_{-1{\bf k}_4} )
+c_0a^\dagger_{0{\bf k}_1}a^\dagger_{0{\bf k}_2}a_{0{\bf k}_3}a_{0{\bf k}_4}\nonumber\\
&\ &+2(c_0+c_2)(a^\dagger_{1{\bf k}_1}a^\dagger_{0{\bf
k}_2}a_{1{\bf k}_3}a_{0{\bf k}_4}+a^\dagger_{-1{\bf
k}_1}a^\dagger_{0{\bf k}_2}a_{-1{\bf k}_3}a_{0{\bf k}_4})
+2(c_0-c_2)a^\dagger_{1{\bf k}_1}a^\dagger_{-1{\bf k}_2}a_{1{\bf k}_3}a_{-1{\bf k}_4}\nonumber\\
&\ &+2c_2(a^\dagger_{1{\bf k}_1}a^\dagger_{-1{\bf k}_2}a_{0{\bf
k}_3}a_{0{\bf k}_4}+a^\dagger_{0{\bf k}_1}a^\dagger_{0{\bf
k}_2}a_{1{\bf k}_3}a_{-1{\bf
k}_4})\bigg{\}}.\label{interactionH} \ea
\end{widetext}
The interaction involves self-scattering and cross-scattering
terms among the three particle flavors, in addition to the last
term which involves the conversion of pairs of $m_F=0$
particles to $m_F=\pm1$ and vice versa.  This provides for
interesting interspecies dynamics that we will explore further
in later sections.

\section{\label{exptBK}An interesting experiment on $^{87}$Rb}

In an interesting experiment carried out by the Stamper-Kurn
group at Berkeley\cite{Sadler:2006}, an initial sample of
$^{87}$Rb atoms was prepared in the $m_F=-1$ state and trapped
in a quasi-two-dimensional optical trap with oscillation
frequencies
$\{\omega_x,\omega_y,\omega_z\}=2\pi\{56,350,4.3\}$s$^{-1}$ at
a longitudinal magnetic field $B_z=2$G.  Subsequently, the
atoms were converted to the $m_F=0$ state by the application of
a radio frequency(r.f.) field, reaching a peak density of
$n=2.8\times 10^{14}$cm$^{-3}$.  The magnetic field was then
quickly ramped down linearly in 5ms to  about 50mG and the gas was
allowed to evolve freely in the trap.  The presence of the
original magnetic field provides a quadratic Zeeman interaction
which lifts the degeneracy for the transitions
$\ket{m_F=-1}\to\ket{m_F=0}$ and $\ket{m_F=0}\to\ket{m_F=1}$
states, so that good conversion may be achieved from the
initial $m_F=-1$ states to $m_F=0$ states. Since $c_2<0$ for $^{87}$Rb, the
interactions are ferromagnetic in nature, and it is thus
energetically favorable for the spins to align with each other.
So, when the magnetic field is ramped down and becomes
effectively zero,  the $m_F=0$ states become dynamically
unstable because of the exchange interaction and convert to
$m_F=\pm1$ pairs.    The different variables in the experiment
are summarized in Table(\ref{BKtable}).  Within some time,
$T_{hold}$, the transverse magnetization in the $xy$-plane was
then imaged, using a novel non-destructive in situ technique,
by detecting its Larmor precession about a guide field. The
experiment revealed that the longitudinal magnetization was
negligible.  In contrast, the images of the transverse
magnetization,  indicated the formation of multiple randomly
oriented domains of varying shapes and sizes, as well as more
involved spin textures.  The typical size, $\xi_{exp}$, of the
domains seen after the domain growth saturated was in the
region of $\sim10\mu$m.  The growth of the transverse
magnetization was observed to be initially exponential, with a
time constant $\tau\sim15(4)$ms.

If one tries to model the initial seeding of these domains
using the Gross-Pitaevskii equations and starting from a pure
$\ket{N}_{m_F=0}$ state, then no evolution of transverse
magnetization is observed\cite{Saito:2007,Yi:2003}. We may
think of the Gross-Pitaevskii equations as corresponding
 to non-linear Schr\"odinger equation describing the motion that begins  sitting on top of a potential hill.
  Without an initial displacement nothing happens classically.  Some form of noise is required to provide the
  instability to roll off down the potential hill.  In a recent paper by Saito et al.\cite{Saito:2007}
  different forms of noise, such as white noise or noise due to the unconverted initial $m_F=-1$ population,
  were tried out to reproduce the instability seen in the experiment by Sadler et al.\cite{Sadler:2006}.
  In this paper we instead model the instability that causes the seeding of the domains observed experimentally
  in terms of quantum noise, and  obtain an analytic form for this in the initial time regimes that the seeding takes place.
  Recent work by Saito et
al.\cite{Saito:2007a} uses a similar picture.
\begin{table}
\center
\begin{tabular}{|l|rl|}
\hline
\multicolumn{3}{|c|}{\emph{$^{87}$Rb data}}\\
\hline
$a_0$ scattering length&101.8&$a_B$\\
$a_2$ scattering length& 100.4&$a_B$\\
Mass &87&g mol$^{-1}$\\
 \hline
\multicolumn{3}{|c|}{\emph{Experimental Parameters}}\\
\hline
Number of atoms&2.1(1)$\times10^6$&\\
Peak density, $n$ & 2.8$\times10^{14}$&cm$^{-3}$\\
Temperature, $T$&40 & nK\\
$c_0n/k_B$ energy scale &$\sim$ 100& nK\\
$c_2n/k_B$ energy scale &$\sim $ 480& pK\\
Trap frequencies $\{\omega_x,\omega_y,\omega_z\}$& 2$\pi\{56,350,4.3\}$&s$^{-1}$\\
Initial magnetic field, $B_z$&2&G\\
Final magnetic field, $B_z$&50&mG\\
Time duration, $T_{hold}$ &36-216&ms\\
$\hbar/(c_2 n)$time scale&$\sim$15.8&ms\\
Observed time constant, $\tau$ &15(4)&ms\\
Typical domain size, $\xi_{exp}$&$\sim$10&$\mu$m\\
\hline
%\multicolumn{3}{|c|}{\emph{Experimental Results}}\\
%\hline
%Typical domain size & $\sim$10& $\mu$m\\
%Instability exponential timescale, $\tau_{fm}$&15(4)&ms\\
%Longitudinal magnetization &negigible&\\
%\hline
\end{tabular}
\caption{Experimental values of interest for an experiment on
$^{87}$Rb \cite{Sadler:2006}.} \label{BKtable}
\end{table}

\section{\label{singleMode}The single mode Hamiltonian and its dynamics}
The simplest approach one can take beyond mean-field theory is
to look at the single mode Hamiltonian.  This has been done by
various authors \cite{Diener:2006,Law:1998}, including some
numerical work within a classical
framework\cite{Robins:2001,Saito:2005, Zhang:2005}. In this
section we will independently reproduce some of the previous
results in a different approach and supplement them with new
insights.

We start with the effective Hamiltonian, for $N$ spin-1 bosons
in zero magnetic field and we assume that all the $m_F$ species
have the same spatial wave function, $\eta({\bf r})$.  This
wave function is determined by using a Gross Pitaevskii
equation which comes from the kinetic part of the Hamiltonian
and the $c_0$ coupling - the symmetric parts of the
Hamiltonian, Eq.(\ref{Hfull}).  Namely, \ba
\left(-\frac{\hbar^2}{2m}\nabla^2+V+c_0N|\eta({\bf
r})|^2\right)\eta({\bf r})=\mu_0\eta({\bf r}), \ea where
$\mu_0$ is a chemical potential enforcing the total particle
conservation.  Since for $^{87}$Rb $c_0\gg c_2$, this means the
spatial wave functions, $\eta({\bf r})$, of the condensate are
largely dependent on $c_0$ and may be taken to be the same for
each $m_F$ species to a first aproximation.   The fields are
thus approximated by \be \psi_{i}({\bf r})=\eta({\bf r}) a_i;\
\psi_{i}({\bf r})^\dagger=\eta({\bf r})^\star a_i^\dagger. \ee
Here we have defined the operators $a_i$, $a_i^\dagger$ that
respectively annihilate or create a boson in the state
$F_z=m_F=i$, with $i\in\{-1,0,1\}$ in the $z$-basis
representation of the spin-1 operators.  These operators obey
the commutation relation \be [a_i,a^\dagger_j]=\delta_{i,j},
\ee with other commutators being zero.  We can further define a
number operator that counts the number of bosons in the state
$i$ as $\hat{N}_i=a_i^\dagger a_i$.   Using these operators,
and taking into account the total number conservation,
$N=\hat{N}_0+\hat{N}_1+\hat{N}_{-1}$, we may rewrite the
Hamiltonian in the form of Diener et al.\cite{Diener:2006}: \ba
H_0 &=&\frac{c}{2}\bigg{(}(\hat{N}_1-\hat{N}_{-1})^2 +(2\hat{N}_0-1)(\hat{N}_1+\hat{N}_{-1})\nonumber\\
&\ &+2a_1^\dagger a_{-1}^\dagger
a_0^2+2a_1a_{-1}(a_0^\dagger)^2\bigg{)}.\label{Hsinglemode} \ea
%:eqn Hsinglemode
c is given by: \be c=c_2\int d^3r|\eta(r)|^4. \ee Notice here
that the $c_0$ terms which would relate to density fluctuations
vanish up to a constant and thus do not affect the dynamics of
the system and do not enter the single-mode Hamiltonian
considered above.

\subsection{\label{DynamicalConsiderations}Dynamical considerations}
We make here the Bogoliubov approximation that we may replace
the annihilation-creation operators for the condensate with
numbers, $a_0\approx a_0^\dagger\approx\sqrt{N_0}$, up to a
phase factor that we may neglect since we are later concerned
only with expectation values where the phase would cancel out.
This approximation is valid for extremely short times as if our
initial state $\ket{N}_0$ condensate acts as an unlimited
source for the creation of particles in the $m_F=\pm1$ states
that does not get depleted.   Also, we notice that in this
approximation, the numbers of particles in the two spin states
$m_F=\pm1$ are equal at all times.  This means we can ignore
the first term in Eq.(\ref{Hsinglemode}) that involes the
difference $(\hat{N_1}-\hat{N}_{-1})$.   Then the Bogoliubov
approximation Hamiltonian becomes quadratic in the fields, \be
H_B=\frac{c}{2}\left((2N_0-1)(\hat{N}_1+\hat{N}_{-1})+2N_0(a_1^\dagger
a_{-1}^\dagger+a_1a_{-1})\right).\label{Hbogoliubov}
%eqn: Hbogoliubov
\ee We may obtain Heisenberg equations of motion in this regime
as follows: \ba
i\hbar\partial_t a_1&=&[a_1,H_0]%=\frac{c}{2}(2 N_0-1)[a_1,a^\dagger_1a_1]+\frac{c}{2}[a_1,2a_1^\dagger a_{-1}^\dagger N_0]
=\frac{c}{2} (2N_0-1)a_1+cN_0a_{-1}^\dagger;\\
i\hbar\partial_t a_{-1}^\dagger&=&
-\frac{c}{2}(2N_0-1)a_{-1}^\dagger-cN_0a_{1}. \ea This system
of linear differential equations can be solved exactly  to
give: \ba
a_1(t)&=&A_1(t)a_1+A_{-1}(t)a_{-1}^\dagger,\\
a_{-1}(t)&=&A_{1}^\star(t)a_{-1}^\dagger+A_{-1}^\star(t)a_1.
\ea where operators with no explicit time dependence correspond
to operators at initial times, $t=0$, and in addition we have
defined,

\ba A_1(t)&\equiv&\cosh \left[\frac{ct \sqrt{4
N_0-1}}{2\hbar }\right]
 \nonumber\\
&\ &-\frac{i\left(2
   N_0-1\right)}{\sqrt{4N_0-1}}\sinh \left[\frac{ct \sqrt{4
   N_0-1}}{2\hbar }\right],\\
A_{-1}(t)&\equiv& -\frac{2i
   N_0}{\sqrt{4N_0-1}} \sinh \left[\frac{ct \sqrt{4
   N_0-1}}{2\hbar }\right].
\ea

\subsection{\label{averages}Number averages and variances}
It is interesting to calculate the expectation value of the
operator $\hat{N}_1(t)$ starting with a ground state where
$N_1=0$. From the results in the previous section and starting
with the  ground state
$\bra{0}_{m_F=-1}\bra{N}_{m_F=0}\bra{0}_{m_F=1}\equiv\bra{0;N;0}$ we get the number expectation values, \ba
N_{\mu}(t)&=&\bra{0;N;0}\hat{N}_\mu(t)\ket{0;N;0}\nonumber\\
&=&\left|\frac{2(N_0)\sinh \left[
\frac{ct\sqrt{4N_0-1}}{2\hbar}\right]}{\sqrt{4N_0-1}}\right|^2,\label{singleModeN}
\ea with $\mu\in\{\pm 1\}$.  This indicates rapid population
transformation from $N_0$ to equal numbers of $m_F=\pm 1$
particles.  The equality of $m_F$  populations is guaranteed
from the Hamiltonian and the conservation of the total number
of particles and the total spin in the system.    We may
investigate the single-mode dynamics further by calculating the
variance,
\begin{widetext}
\ba
\sigma_{\mu}^2&=&\overline{N_{\mu}(t)^2}-\left(\overline{N_{\mu}(t)}\right)^2=|A_1|^2|A_{-1}|^2\\
&=&\frac{4 \sinh ^2\left[\frac{c t \sqrt{4 N_0-1}}{2\hbar
}\right] N_0^2 \left(4 N_0
   \left(N_0 \sinh ^2\left[\frac{c t \sqrt{4 N_0-1}}{2\hbar
   }\right]+1\right)-1\right)}{\left(1-4 N_0\right)^2}\\
 &=&\overline{N_{\mu}(t)}\left(\overline{N_{\mu}(t)}+1\right).
\ea
\end{widetext}
This is exactly the variance of a thermal
state\cite{Lifshitz:1984a} and it actually corresponds to a
Bose-Einstein distribution - we will see this explicitly in the
context of our multiple mode treatment in
Sec.(\ref{effectiveAlgebra}).  The single mode condensate
starts out with a source of $N$ particles in the $m_F=0$ pure
state that are transmuted into equal numbers of $m_F=\pm 1$.
Each of the resulting species displays a super-Poissonian
variance, having a temperature and entropy associated with it.
We should keep in mind that this is done in the Bogoliubov
approximation and is in need of further exploration.  Such
considerations are undertaken in the next section for exact
single mode solutions for various numbers of particles, as well
as in Sec.(\ref{multimode}), which explores a multi-mode
approximation.

\subsection{\label{exactN}Exact single mode solutions for different numbers of particles}
\begin{figure}[t]
\includegraphics[width=3.375in]{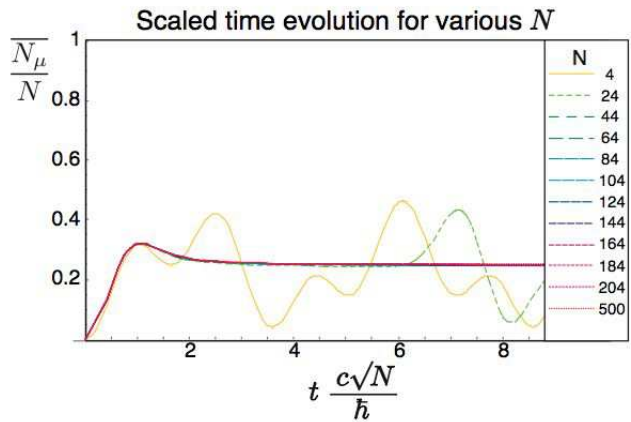}
\caption{\label{timeEvolutionN}(Color online) The single mode Hamiltonian
may be solved exactly for different $N$.  The time evolution of the solutions indicates scaling of
$t \sim \frac{1}{\sqrt{N}}$ as the curves collapse into a single line for short times.}
\end{figure}
%%
%:figure:timeEvolutionNAll
\begin{figure}[t]
\includegraphics[width=3.375in]{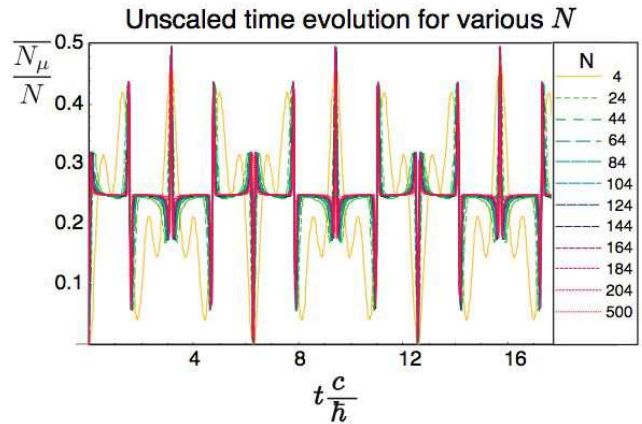}
\caption{\label{timeEvolutionNAll}(Color online) The exact solutions for
the single mode Hamiltonian exhibit a linear time regime, with the curves for different $N$ collapsing
into a single curve without scaling.  The populations oscillate and ultimately complete many cycles of
 non-linear evolution followed by linear evolution.}
\end{figure}

We now turn our attention to solving the effective single mode Hamiltonian, Eq. (\ref{Hsinglemode})  exactly for
different initial numbers of particles $N$. Given
our starting state $\ket{N}_{m_F=0}$, only pairs of particles
may be created or destroyed, and this makes for a smaller
subspace under our consideration.  If we take our initial state
to be the `number of pairs' vacuum state, for $N$ particles we
may only have $\frac{N}{2}+1$ number of pair-states.  We may
write the matrix components of the Hamiltonian that connects
pair-states with $i$ and $ j$ pairs respectively as:
\ba
\tilde{H}_{ij}&\equiv&\frac{c}{2}\bigg{\{}2i\left(2\left(N-2i\right)-1\right)\delta_{i,j}\nonumber\\
&\ &+2\sqrt{(N-2j)(N-2j-1)}\delta_{i,j+1}\nonumber\\
&\ &+2\sqrt{(N-2j+1)(N-2j+2)}\delta_{i,j-1}\bigg{\}},
\ea
where $0\leq i,j\leq N/2$. We may think of this as a hopping
Hamiltonian, allowing for movement between adjacent states that
differ by one pair.  This is essentially a one-dimensional
problem, resulting in a tridiagonal matrix that may be solved
numerically to obtain the eigenvalues and time evolution of the
system.  The time evolution for various N is shown in the two
figures, Fig.({\ref{timeEvolutionN}) and
Fig.(\ref{timeEvolutionNAll}).

%:figure:ComparisonBogoliubov204
\begin{figure}[b]
\includegraphics[width=3.375in]{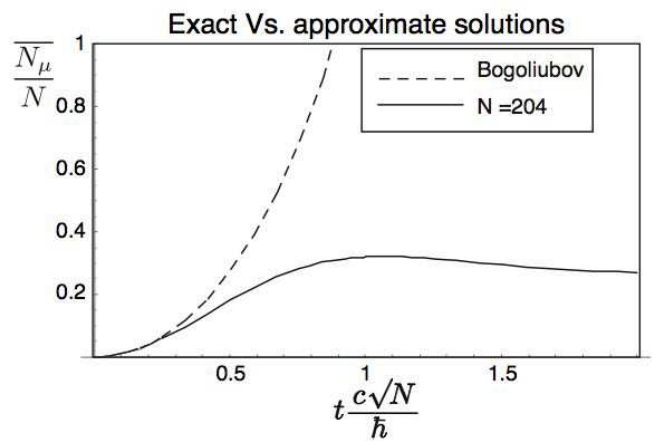}
\caption{\label{ComparisonBogoliubov204}The exact solution for $N=204$
is compared to the Bogoliubov approximation.  As is seen from the graph the two are in good agreement at
the very early initial times, in support of our no-depletion approximation for the single mode Hamiltonian.}
\end{figure}

%:figure:varianceScaledN
\begin{figure}[t]
\includegraphics[width=3.375in]{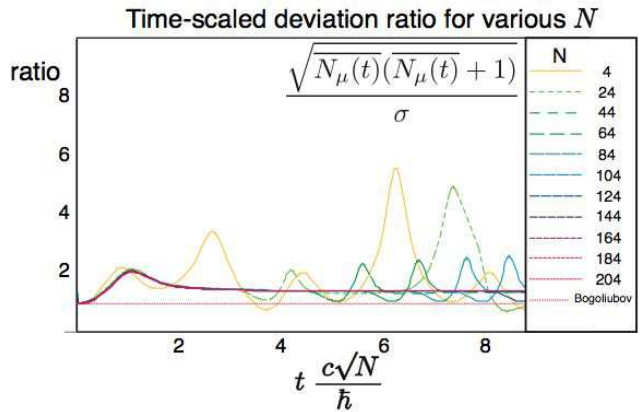}
\caption{\label{varianceScaledN}
(Color online) The ratio $\sqrt{\overline{N_\mu(t)}(\overline{N_\mu(t)}+1)}$ to the deviation $\sigma$  is a measure
of the correlation of the state variance to that of  a thermal state.  Our Bogoliubov approximation is
shown to obey a thermal variance and to match the exact results only for very short times.  The various
 curves can be seen to collapse again into a single curve showing that the plotted ratio is independent of $N$.}
\end{figure}

As we can see from the figures, we may identify a critical time
$t_c\sim \frac{1}{\sqrt{N}}$ which separates two time regimes,
in agreement with the results of Law et al\cite{Law:1998}.  We
also notice the additional feature of scaling with
$\frac{1}{\sqrt{N}}$ for short time scales, where the different
evolutions for different total number of particles collapse
into a single curve as seen in Fig.(\ref{timeEvolutionN}).
There are two time regimes in the system:  First a regime
within which time scales as $\frac{1}{\sqrt{N}}$ when the
dynamics is dominated by the non-linear part of the Hamiltonian
shown in Fig.(\ref{timeEvolutionN}).  Then we see a subsequent
metastable time regime, when the linear part of the Hamiltonian
takes over the time evolution and the time does not scale with
the number of particles, Fig.(\ref{timeEvolutionNAll}).  The
time regimes alternate for many cycles if the system is allowed
to evolve over time.  The starting condensate of $N$ particles
in the $m_F=0$ state is initially rapidly depleted to create
pairs of $m_F=\pm1$ particles.  When the metastable time regime
is reached, at $t\frac{c\sqrt{N}}{\hbar}=1$,  the population
ratios are
\be
\frac{N_{-1}}{N}:\frac{N_{1}}{N}:\frac{N_{0}}{N}\approx
0.25:0.25:0.5.
\ee
This metastable regime is followed by oscillations in the three different populations and eventually the complete repopulation of the $m_F=0$ state, reaching again the initial state and the cycle repeating itself when $t\frac{c}{\hbar}=2\pi$.

Furthermore, we may compare our Bogoliubov transformation for
short times to the exact solution for a given $N$. As shown in
Fig.(\ref{ComparisonBogoliubov204}) the two are in agreement
only for very short times.  Assuming the scaling of time with
$\frac{1}{\sqrt{N}}$ this implies a validity time  dictated by
the depletion, where the percentage, $\kappa_0$, of particles
remaining in the $\ket{N}_{m_F=0}$ state  is on the order of
$\kappa_0=\frac{N_0-\sum_{\mu=\pm1}\_{m_F=0}\bra{N}\hat{N}_{\mu}\ket{N}_{m_F=0}}{N_0}\sim0.9$.  We are interested in the seeding of the domains at very early times in the time-evolution and hence the Bogoliubov
approximation looks indeed like a good starting point.

Additionally, the statistical variance and deviation were
investigated for various functional forms of the number
expectation, $\overline{N_\mu(t)}$.  It was found that the
curves for different number of particles collapse into single
curves when we scale the variance with $\overline{N_\mu(t)}(\overline{N_\mu(t)}+1)$.  Plots for
different time scalings corresponding to the two time regimes
identified above are shown in Fig.(\ref{varianceScaledN}) and
Fig.(\ref{varianceN}).   In the plots the ratio of
$\sqrt{\overline{N_\mu(t)}(\overline{N_\mu(t)}+1)}$ to the
deviation $\sigma$, reveals scaling and collapse of the plotted
curves showing that the deviation must be intimately related to
a thermal distribution.  The results of the Bogoliubov
approximation are included for reference.  As shown, the
Bogoliubov approximation obeys a thermal variance through all
time regimes, displaying super-Poissonian statistics consistent
with two-mode squeezing
picture\cite{Walls:1983,Walls:1995,Puri:2001,Caves:1985}, which
will be discussed further in terms of the multimode
approximation in the next section.

%:figure:varianceN
\begin{figure}[t]
\includegraphics[width=3.375in]{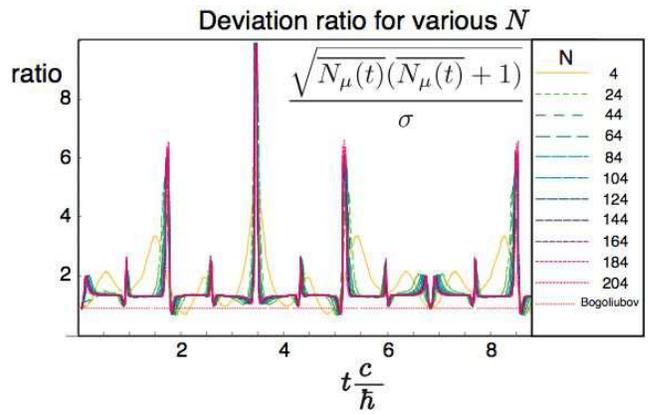}
\caption{\label{varianceN}(Color online) The ratio $\sqrt{\overline{N_\mu(t)}(\overline{N_\mu(t)}+1)}$ to the deviation $\sigma$  is here shown for long times.  We still observe that
 our Bogoliubov approximation always obeys a thermal variance.  Furthermore, the plot indicates that the
  plotted ratio is independent of $N$ indicating that the actual variance is some functional of a thermal variance.}
\end{figure}
%section: multimode
\section{\label{multimode}Multimode Bogoliubov approximation}
We now would like to extend our discussion to the case of many
spatial modes.  The picture we have in mind is that $m_{F}=\pm1$ pairs are created from the $m_F=0$ source and then separate
to begin seeding of the domains.  The interplay of the various
wavevectors ${\bf k}$ will then be important in ascertaining
the real space distribution and behavior of these domain seeds.
We first look at an effective Hamiltonian description.
\subsection{\label{approx0}The effective multimode Hamiltonian}
We begin with Eq.(\ref{interactionH}) and make the
approximations that:  we have no  $m_F=0$ atoms in a
non-condensed state; self scattering and cross scattering terms
outside of the spinor $m_F=0$ condensate maybe be ignored; we
may take our Bogoliubov approximation once more for the $m_F=0$
operators, $a_0\sim a_0^\dagger\sim\sqrt{N}$. Then the
effective Hamiltonian becomes, \ba
H_{IB}&=&\frac{c_0}{2V}N^2+N \sum_{{\bf k}}\bigg{\{}\frac{(c_0+c_2)}{V}
(a^\dagger_{1{\bf k}}a_{1{\bf k}}+a^\dagger_{-1{\bf k}}a_{-1{\bf k}})\nonumber\\
&\ &+\frac{c_2}{V}(a^\dagger_{1{\bf k}}a^\dagger_{-1-{\bf
k}}+a_{1{\bf k}}a_{-1-{\bf k}})\bigg{\}}. \ea

As a simplification we will be assuming  we can drop the $c_0$
terms in some short-time `no-depletion' approximation since the
variation of the $c_0$ terms with respect to $N_0$ approaches
zero in this approximation - as we recall the terms cancel out
identically in the single-mode approximation.  The magnitude of
the coupling ratio $\frac{c_0}{c_2}\sim 10^4 $ suggests that
density fluctuation modes, as dictated  by the $c_0$ terms, are
very stiff, so they are ignored in our calculations.  We
further neglect the spatial dependence of the trap potential.
With all these simplifications made, we can finally write down
the effective Hamiltonian we will use for our multimode
treatment of the spinor condensate,
\ba
H_{E}&=&\sum_{{\bf k}}\bigg{\{}(\epsilon_{\bf k}+c_2
n)(a^\dagger_{1{\bf k}}a_{1{\bf k}}+a^\dagger_{-1{\bf
k}}a_{-1{\bf k}})\nonumber\\
&\ &+c_2 n(a^\dagger_{1{\bf k}}a^\dagger_{-1-{\bf
k}}+a_{1{\bf k}}a_{-1-{\bf k}})\bigg{\}},
\label{Heffmultimode}\ea

%:eqn Heffmultimode
where we have defined the kinetic energy, 
\be \epsilon_{\bf
k}\equiv\frac{\hbar {\bf k}^2}{2 m}.
\ee
The effective Hamiltonian above treats the system in terms of pairs of the
two species with $m_F=\pm1$ having opposite wavevectors.  The
first term in the Hamiltonian counts the number of pairs and
the second term provides an interaction via creation and
annihilation of the said pairs.  The Hamiltonian is quadratic
in the field operators and thus we are looking  fluctuations of
the system at a gaussian level.
\subsection{\label{TimeEvolution}Time evolution of operators, noise and statistics}
For a given mode we may obtain the following equations of
motion,
\ba
i\hbar \frac{\partial a_{1\bf k}(t)}{\partial t}&=& \left(\epsilon_{\bf k}+c_2 n \right) a_{1 \bf k}(t)+c_2 n a_{-1-\bf k}^\dagger(t);\label{a1kmotion}\nonumber\\
%:eqn a1kmotion
-i\hbar \frac{\partial a_{-1\bf -k}^\dagger(t)}{\partial t}&=&\left(\epsilon_{\bf k}+c_2 n \right) a_{-1- \bf
k}^\dagger(t)+c_2 n a_{1\bf k}(t).\label{amin1kmotion}\nonumber\\
\ea
%:eqn amin1kmotion
As for the single mode case these may be solved exactly to
obtain \ba
a_{1\bf k}(t)&=&A_{1 \bf k}(t)a_{1\bf k}+A_{-1-\bf k}(t)a_{-1-\bf k}^\dagger;\nonumber\\
a_{-1-\bf k}^\dagger(t)&=&A_{1 \bf k}^\star(t)a_{-1-\bf
k}^\dagger+A_{-1-\bf k}^\star(t) a_{1\bf k}, \ea where we now have
defined \ba
A_{1 \bf k}(t)&\equiv& \cosh\left[\frac{t}{\hbar}G_{\bf k}\right]-\frac{i( c_2  n
+\epsilon_{\bf k})}{G_{\bf k}}\sinh\left[\frac{t}{\hbar}G_{\bf k}\right]; \nonumber\\
A_{-1\bf k}(t) &\equiv & \frac{- i c_2  n}{G_{\bf
k}}\sinh\left[\frac{t}{\hbar}G_{\bf k}\right], \ea with an
effective `gain'  function for a given mode ${\bf k}$, \be
G_{\bf k}^2\equiv \epsilon_{\bf k}(-2 c_2  n - \epsilon_{\bf
k}). \ee Again, in these solutions operators with no explicit
time dependence refer to the inital time, $t=0$, operators.
The paper by Saito et al.\cite{Saito:2007a} also arrives at the
same solutions to the differential equations, with a similar
treatment of ignoring the $c_0$ terms. The current work
supplements these with single mode considerations, and with an
alternative treatment that is developed below: a quantum noise
approach where a new connection is made to the
$\mathfrak{su}(1,1)$ language of quantum squeezing.

The two operators $a_{1 \bf k},~a_{-1-\bf k}$ are closely
intertwined, with the population of one acting as a noise
source for the other and vice versa.  This suggests also that
the growth of different species with opposite wavevectors is a
highly correlated process. Indeed, the correlation of species
of opposite wavevectors  is non-zero,
\ba
&&_{m_F=0}\bra{N}a_{-1{\bf k}'}(t)a_{1\bf k}(t)\ket{N}_{m_F=0}=\nonumber\\
&=&\bigg{\{}\frac{c_2n(c_2n+\epsilon_{\bf k})}{G_{\bf k}}\sinh\left[\frac{t}{\hbar}G_{\bf k}\right]^2\nonumber\\
&\ &-\frac{ i c_2  n}{G_{\bf
k}}\sinh\left[\frac{t}{\hbar}G_{\bf
k}\right]\cosh\left[\frac{t}{\hbar}G_{\bf
k}\right]\bigg{\}}\delta_{{\bf k},{-\bf k'}},
\ea where we may
picture that the creation of a $\ket{1}_{1 \bf {k}}$ state is
correlated with the creation of a $\ket{1}_{-1-\bf k}$ state.
The $N$-particle condensate source acts as a reservoir for the
creation of pairs of particles moving in opposite directions.
The $m_F=\pm1$ states are populated rapidly, at an exponential
rate, as shown by the evolution of the number expectation for a
given wavevector: \ba
\overline{N_{\mu\bf k}(t)}&=&_{m_F=0}\bra{N}\hat{N}_{\pm 1\pm\bf k}\ket{N}_{m_F=0}\nonumber\\
&=&\frac{c_2^2 n^2}{G_{\bf k}^2}\sinh\left[\frac{t}{\hbar}G_{\bf k}\right]^2.\label{Nkavg}\nonumber\\
%:eq:Nkavg
\ea We should note here that \be \lim_{t\rightarrow
0}\overline{N_{\mu\bf k}(t)}=\frac{c_2^2 n^2 t^2}{\hbar^2}, \ee
which agrees with our single mode considerations in the same
limit, Eq.(\ref{singleModeN}).

%:figure:kvectors
\begin{figure}[tb]
\includegraphics[width=3.375in]{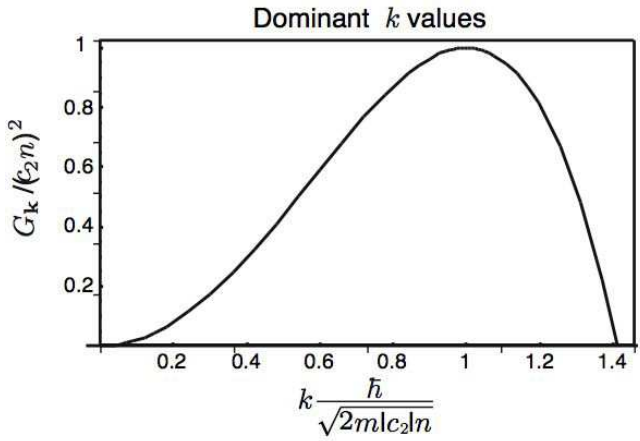}
\caption[Dominant $k$ values]{\label{kvectors} The gain parameter $G_{\bf k}$ is real
for only a set of values of $k$ and these are the major contributors in the exponential
 growth of the number of $m_F=\pm 1$ particles.  Imaginary values lead to oscillatory behavior
  instead which does not support the number growth of the two species.}
\end{figure}

The exponential growth of the  number of particles with time
implies  our no-depletion approximation is only valid in the
initial stages of the time evolution.  Furthermore, the growth
is dominated by certain values of the magnitude of the
wavevectors ${\bf k}$ as seen by the behavior of $G_{\bf k}^2$,
Fig.(\ref{kvectors}), and is independent of direction.    The
dominant wavector is at the maximum of $G_{\bf k}^2$: \ba
k_{max}= \sqrt{\frac{-2 m c_2 n}{\hbar^2}}. \ea

%:figure:Number Evolution
\begin{figure}[tb]
\includegraphics[width=3.375in]{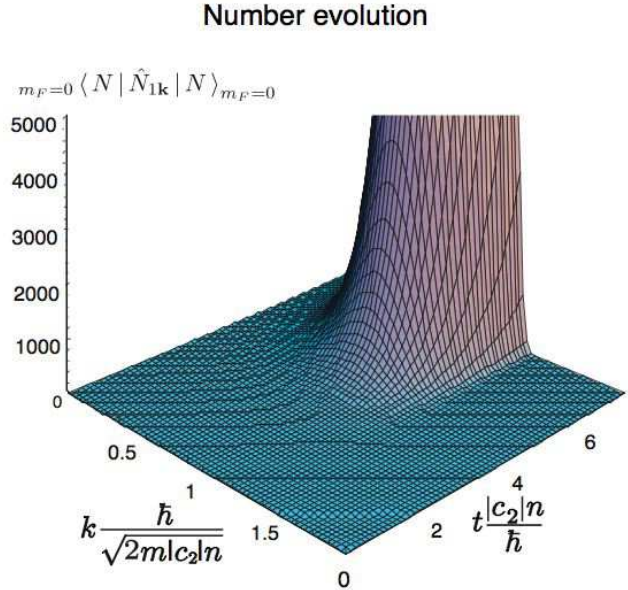}
\caption{\label{Nevolution}(Color online) For $^{87}$Rb, the expectation number of particles
for a given species grows over a certain range of values of the wavevector magnitude $k$.  Furthermore
the growth in time is exponential and so our no-depletion approximation remains valid only at earlier times }
\end{figure}

This is the wave vector with the greatest initial instability.
For $^{87}$Rb, $c_2<0$ and hence $k_{max}\in \mathbf{R}$. It
defines a coherence length scale, \be \xi_{0}\sim
\frac{1}{k_{max}}=\sqrt{-\frac{\hbar^2}{2mc_2 n}}\approx
2.4\mathrm{\mu m}, \ee which in turn gives a coherence volume
\be
 V_{\xi_0}\sim \frac{4\pi\xi_{0}^3}{3},
\ee which for the experimental peak density,
$n=2.8\times10^{-14}$cm$^{-3}$, has associated with it $N_{\xi}=n V_{\xi_0} \sim 1.6\times10^4$ particles.  This is
large enough to support our assumption that the initial
fluctuations are gaussian in nature, as suggested in our
approximation of using a quadratic Hamiltonian.  This allows us
to explore the small quantum fluctuations that provide the
starting instability for the evolution of the system - assuming
no depletion of our initial particle source.  We may
investigate further by looking at the number evolution in time
as a function of wavevector as shown in Fig.
(\ref{Nevolution}). This again shows the range of
wavevectors over which the instability causes the growth of the
$m_F=\pm 1$ states.  Additionally, we observe that the growth
is exponential in time, and that initially our depletion
approximation should be valid.  As an aside, we note here that
for $^{23}$Na $G_{\bf k}$ does not exhibit a real $k_{max}$
since for this system $c_2>$0. Hence for $^{23}$Na the number
expectation values are oscillatory and no strong evolution into
$m_F=\pm 1$ species would take place, consistent with the
`antiferromagnetic' interactions in this system\cite{Black:2007}.
%:figure:depletion
\begin{figure}[tb]
\includegraphics[width=3.375in]{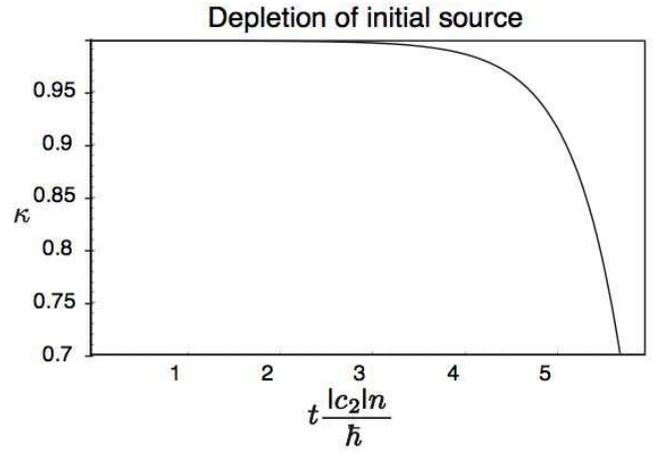}
\caption{\label{depletion} The population of the $m_F=\pm1$ species leads
to the depletion of the initial source of $N_0$ atoms in the $m_F=0$ state.  Since our approximation
 assumes no depletion we can only ascertain its validity in the initial time region in the figure where it
 is only a few percent of the total available $N_0$ reservoir.}
\end{figure}
We may calculate the percentage of particles remaining in the
$\ket{N}_{m_F=0}$ state source of the system as a function of
time.     We have, \ba \kappa=\frac{N_0-\sum_{\mu,{\bf k}}\
_{m_F=0}\bra{N}\hat{N}_{\mu \bf k}\ket{N}_{m_F=0}}{N_0}. \ea
The sum may not be calculated in closed form, but may be
approximated by integrating the wavevector over the
``instability'' range.  The results are shown in
Fig.(\ref{depletion}), again indicating that at early times the
depletion is only a few percent and our approximations must
still be valid. For the characteristic time $t\frac{|c_2|n}{\hbar}$ we can substitute for the Berkeley experiment parameters to obtain each time unit
as $\sim16$ms.  Then up to say $4\times16$ms in
Fig.(\ref{depletion}) corresponds to the first two initial
images from the experiment, as shown in Fig.(2) in the paper by
Sadler et al.\cite{Sadler:2006}. The experimental results are
consistent with our no-depletion approximation for the initial
times, when the domains are still forming and the transverse
magnetization has not yet saturated as seen from the brightness
of the figures.

%subsection
\subsubsection{Thermal Statistics}
Having calculated the number expectation for a given species
and mode it is natural to consider its variance.  It turns out
that we may express our result simply, in terms of the number
operator expectation value in the $m_F=0$ starting state
evolving in time, 
\ba
\sigma_{\mu\bf k}^2&=&\left<N_{\mu {\bf k}}(t)^2\right>-\left<N_{\mu {\bf k}}(t)\right>^2\nonumber\\
&=&\frac{ c_2^2 n^2 \left(G_{\bf k}^2+ c_2^2 n^2\sinh ^2\left[\frac{t  }{\hbar }G_{\bf k}\right]\right)
\sinh ^2\left[\frac{t  }{\hbar }G_{\bf k}\right]}{G_{\bf k}^4}\nonumber\\
&=&\left<N_{\mu {\bf k}}(t)\right>\left(\left<N_{\mu {\bf
k}}(t)\right>+1\right).
\ea
This is exactly the same form as for a thermal state.  This may be slightly counterintuitive because we start out with a pure state at zero temperature,
with no entropy assocciated with it and end up with essentially
a thermal distribution (albeit with a different temperature for
each spatial $k$ mode). If we were to imagine that we were aware of the existence of only one of the two species, and carried a measurement of only
that species for a given wavevector, we would associate a temperature to it based on its variance. Equivalently, we may think of this as what we would get if say we traced out the $m_F$=-1 atoms: the $m_F$=1 atoms display an effective finite entropy and a thermal distribution. These super-Poissonian
statistics show atom bunching and bear similarities to processes in as diverse fields as quantum optics \cite{Gardiner:2000,Puri:2001,Walls:1983,Walls:1995}, topological defects and particle production in cosmology and gravitation\cite{Damski:2007,Guth:1985,Hamilton:2004, Hawking:1975,Kibble:1976,Zurek:1985,Saito:2007},
within the common language of squeezing and the su(1,1) algebra
with which we will recast our results in the following section.
Furthermore, we will see how the thermal variance corresponds
to a Bose-Einstein distribution using a reduced density matrix
computation.

%subsection
\subsection{\label{effectiveAlgebra}The $\mathfrak{su}$(1,1) underlying algebra and evolution}
We notice that our Hamiltonian $H_E$ has an underlying
structure in terms of pairs of different species and opposite
wavevectors.  Namely, we define operators that annihilate and
create pairs at a given wavevector ${\bf k}$, $K^-_{\bf k}$ and
$K^+_{\bf k}$ respectively, as well as an operator counting
pairs, $K^0_{\bf k}$:
\ba
K^-_{\bf k}&\equiv& a_{1{\bf k}} a_{-1{-\bf k}},\\
K^+_{\bf k}&\equiv& a_{1{\bf k}}^\dagger a_{-1{-\bf k}}^\dagger,\\
K^0_{\bf k}&\equiv& \frac{1}{2}\left( a_{1{\bf k}}^\dagger
a_{1{\bf k}} +  a_{-1{-\bf k}} a_{-1{-\bf k}}^\dagger\right).
\ea
The commutators of these operators turn out to be:
\ba
\left[ K^0_{\bf k}, K^\pm_{\bf k}\right]&=&\pm K^\pm_{\bf k},\\
\left[ K^+_{\bf k}, K^-_{\bf k}\right]&=&-2 K^0_{\bf k},
\ea
providing a realization of the generators of an
$\mathfrak{su}$(1,1) algebra\cite{Hall:2003,Knapp:2002}.  In
terms of these we may rewrite our effective Hamiltonian as, \be
H_E  = \sum_{{\bf k}}\bigg{\{}\left(\epsilon_{\bf k}+ c_2 n
\right)(2K^0_{\bf k}-1)+n c_2(K^+_{\bf k}+K^-_{\bf
k})\bigg{\}}. \ee We observe that the operators $K_{\bf k}^\pm$
create and annihilate pairs of atoms of different $m_F$
species, with opposite wavevectors.  Then $K_{\bf k}^0$ simply
counts the number of these pairs in the system.   We have thus
 arrived at the $\mathfrak{su}$(1,1) algebra description
that is associated with quantum squeezing.  This is a multimode
realization of the algebra, one realization per pair of atoms
of different species and oppositely directed wavevectors.  As
we have already seen in the previous section the associated
quantum states are characterized by thermal, super-Poissonian
fluctuations, as was already discovered in our single-mode
considerations.

Having obtained the form of the Hamiltonian we may now use it
for unitary time evolution to arrive after time $t$ to some
final state $\ket{f}$,
\begin{widetext}
\ba
\left|\, f\, \right\rangle&=&\exp\left[\frac{-i H_E t}{\hbar}\right]\left|\, N\, \right\rangle_{m_F=0}\\
&=&\prod_{\bf k}\exp[i\phi_{\bf k}]\exp\left[\frac{-i t}{\hbar}\bigg{\{}\left(\epsilon_{\bf k}+ c_2 n \right)2K^0_{\bf k}+n c_2(K^+_{\bf k}+K^-_{\bf k})\bigg{\}}\right]\ket{N}_{m_F=0},\nonumber\\
\ea where, \be \phi_{\bf k}\equiv
\frac{t}{\hbar}\left(\epsilon_{\bf k}+c_2 n \right). \ee We
want to disentangle the exponential as: \be
\exp[\theta\{a^+_{\bf k}\hat{K}^+_{\bf k}+a^0_{\bf
k}\hat{K}_{\bf k}^0+a_{\bf k}^-\hat{K}_{\bf
k}^-\}]=\exp[\phi_{\bf k}^+(\theta)\hat{K}_{\bf
k}^+]\exp[\phi_{\bf k}^0(\theta)\hat{K}_{\bf
k}^0]\exp[\phi_{\bf k}^-(\theta)\hat{K}_{\bf k}^-], \ee
\end{widetext}
where $\theta$ is an auxiliary parameter which is set to one at
the end and we have defined: \ba
a^+_{\bf k}=a^-_{\bf k}&\equiv& \frac{-i t}{\hbar}n c_2%\in i\mathbf{R}
,\\
a^0_{\bf k} &\equiv&2 \frac{-i t}{\hbar}\left(\epsilon_{\bf k}+nc_2\right)%\in i\mathbf{R}
.\ea Matching the coefficients of the generators on both sides
would give \cite{Gerry:1985,Puri:2001}: \ba
a_{\bf k}^+&=&\dot{\phi}_{\bf k}^+-\phi_{\bf k}^+\dot{\phi}_{\bf k}^0+(\phi_{\bf k}^+)^2\dot{\phi}_{\bf k}^-\exp[-\phi_{\bf k}^0],\\
a_{\bf k}^0&=&\dot{\phi}_{\bf k}^0-2\phi_{\bf k}^+\dot{\phi}_{\bf k}^-\exp[-\phi_{\bf k}^0],\\
a_{\bf k}^-&=&\dot{\phi}_{\bf k}^-\exp[-\phi_{\bf k}^0]. \ea
The last two equations may be substituted into the first to
obtain a Ricatti equation\cite{Boyce:2005}, \be \dot{\phi}_{\bf
k}^+-(\phi_{\bf k}^+)^2a_{\bf k}^- -\phi_{\bf k}^+a_{\bf
k}^0-a_{\bf k}^+=0, \ee which gives the solutions: \ba
\phi_{\bf k}^+(\theta)&=&\frac{a_{\bf k}^+}{\Gamma_{\bf k}}\frac{\sinh(\Gamma_{\bf k}\theta)}{\cosh(\Gamma_{\bf k}\theta)-\frac{a_{\bf k}^0\sinh(\Gamma_{\bf k}\theta)}{2\Gamma_{\bf k}}},\\
\phi_{\bf k}^0(\theta)&=&-2\ln[\cosh(\Gamma_{\bf k}\theta)-\frac{a_{\bf k}^0}{2\Gamma_{\bf k}}\sinh(\Gamma_{\bf k}\theta)],\\
\phi_{\bf k}^-(\theta)&=&\frac{a_{\bf k}^-}{\Gamma_{\bf
k}}\frac{\sinh(\Gamma_{\bf k}\theta)}{\cosh(\Gamma_{\bf
k}\theta)-\frac{a_{\bf k}^0\sinh(\Gamma_{\bf
k}\theta)}{2\Gamma_{\bf k}}}, \ea where \ba
\Gamma_{\bf k}^2&=&\frac{(a_{\bf k}^0)^2}{4}-a_{\bf k}^+a_{\bf k}^-\\
&=&\frac{t^2}{\hbar^2}G_{\bf k}^2. \ea Our final state becomes:
\begin{widetext}
\ba \ket{f(t)}&=&\otimes_{\bf k}\left\{
\exp\left[\frac{\phi_{\bf k}^0}{2}+i\phi_{\bf k}\right]\sum_n
(\phi_{\bf k}^+)^n \left|\, n\, \right\rangle_{1 \bf k}
\left|\, n\, \right\rangle_{-1 -\bf
k}\right\}\equiv\otimes_{\bf k}\ket{f_{\bf k}(t)}. \ea
\end{widetext}
This is an exact solution for the dynamics of our effective
Hamiltonian, $H_E$, in terms of number states for a given
wavevector and particle species.   We can either use the
time-evolved state or simply the time-evolution of the creation
annihilation operators to investigate the statistical behavior
of the system and arrive at the familiar expression for
two-mode squeezing \be \left<N_{\mu {\bf
k}}(t)^2\right>-\left<N_{\mu {\bf k}}(t)^2\right>=\left<N_{\mu
{\bf k}}(t)\right>\left(\left<N_{\mu {\bf
k}}(t)\right>+1\right), \ee once again indicating that the
variance for a given mode is thermal in nature.

Let us investigate the thermal variance a bit further.  Say we
look at the density operator $\hat{\rho}_{\bf k}$ for a given
wavevector, which for a pure state corresponds to \ba
\hat{\rho}_{\bf k}&=&\ket{f_{\bf k}(t)}\bra{f_{\bf k}(t)}. \ea
We now perform a partial trace,  to obtain the density matrix
for the $m_F=1$ particles after tracing out the $m_F=-1$
degrees of freedom for a given wavevector ${\bf k}$,
\begin{widetext}
	\ba
\hat{\rho}_{+1\bf k}&=&\mathrm{Tr_{m_F=-1}}\hat{\rho}_{\bf k}\nonumber\\
&=&\frac{1}{\cosh\left[\frac{G_{\bf k}t}{\hbar}\right]^2+\frac{(\epsilon_{\bf k}+nc_2)^2}{G_{\bf k}^2}\sinh\left[\frac{G_{\bf k} t}{\hbar}\right]^2}\sum_{i}\left(\frac{n^2c_2^2\sinh\left[\frac{G_{\bf k}t}{\hbar}\right]^2}{G_{\bf k}^2\left(\cosh\left[\frac{G_{\bf k}t}{\hbar}\right]^2+\frac{(\epsilon_{\bf k}+nc_2)^2}{G_{\bf k}^2}\sinh\left[\frac{G_{\bf k} t}{\hbar}\right]^2\right)}\right)^i\left|\, i\, \right\rangle_{1 \bf k} \bra{i}_{1\bf k}\nonumber\\
&=&\frac{1}{1+\overline{N_{1\bf
k}(t)}}\sum_i\left(\frac{\overline{N_{1\bf
k}(t)}}{1+\overline{N_{1\bf k}(t)}}\right)^i\ket{i}_{1\bf
k}\bra{i}_{1\bf k}. \ea An identical calculation but instead
tracing out the $m_F=1$ states gives, \ba \hat{\rho}_{-1\bf k}
&=&\frac{1}{1+\overline{N_{-1-\bf
k}(t)}}\sum_i\left(\frac{\overline{N_{-1-\bf
k}(t)}}{1+\overline{N_{-1-\bf k}(t)}}\right)^i\ket{i}_{-1-\bf
k}\bra{i}_{-1-\bf k}. \ea
\end{widetext}
The reduced density operators, $\hat{\rho}_{\pm1\bf k}$, have
exactly the form of the density matrix for a Bose-Einstein
distribution\cite{Drummond:2004, Gardiner:2000}. This is
remarkable because our pair state $\ket{f(t)}$ is a pure state
that has no entropy associated with it.  On the other hand the
reduced density operators, $\rho_{\pm 1\bf k}$, correspond to
thermal Bose-Einstein distributions, that have an associated
entropy and noise that results from the perfect correlations
between the $m_F=\pm1$ particles in each pair state. For a
given wavevector, the $m_F=+1$ particles act as noise for the
$m_F=-1$ particles and vice versa and hence production of pairs
from our starting `vacuum' provides a thermalization mechanism
for a given particle species.  The density operator depends on
the wavevector ${\bf k}$ indicating that the different modes of
the system would receive different amounts of thermalization,
and thus no single temperature may be associated to one or the
other species for the entire system.
We may use the above results to produce simulations of the
system similar to those of Saito et al.\cite{Saito:2007a}.
Saito et al. use the classical Gross-Pitaevskii equations but
build in white noise to represent the quantum fluctuations. The
white noise is chosen to reproduce the variance found in the
fully quantum equations of motion which we use.  Here we follow
the
 the $Q$-function method of quantum optics.  The $Q$-function of a density operator
$\rho$ is defined as \cite{Gardiner:2000} \be Q(\alpha,
\alpha^\star)\equiv\frac{1}{\pi}\bra{\alpha}\rho\ket{\alpha}.
\ee For a pure state this is proportional to the probability of
the state to be in a coherent state $\alpha$.  For our
representation of state $\ket{f(t)}$ we may write:
\begin{widetext}
\ba
Q&\equiv& \frac{1}{\pi^{2\eta}}\left| \otimes_{\bf k}\left(\bra{\alpha_{1\bf k}}\otimes\bra{\alpha_{-1-\bf k}}\right)\ket{f(t)}\right|^2\\
&=&\prod_{\bf k}\left(\frac{1}{\pi^{2}}\exp\left[ \frac{\phi_{\bf k}^0+\phi_{\bf k}^{0\star}}{2}-|\alpha_{1\bf k}|^2-|\alpha_{-1-\bf k}|^2
+\phi_{\bf k}^+\alpha_{1\bf k}^\star\alpha_{-1-\bf k}^\star+\phi_{\bf k}^{+\star}\alpha_{1\bf k}\alpha_{-1-\bf k} \right]\right) \nonumber\\
\\&\equiv&\prod_{\bf k}Q_{\bf k},
\ea
\end{widetext}
where $\eta$ is the number of modes and  we have used the
expansion of coherent states in terms of number states, \be
\ket{\alpha}=\exp\left[-\frac{|\alpha|^2}{2}\right]\sum_l\frac{\alpha^l}{\sqrt{l!}}\ket{l}.
\ee Given this probability distribution we notice that each
$Q_{\bf k}$ is an exponential  of a quadratic form and is thus
gaussian in nature.  A change of basis to diagonal form allows
us to calculate the variance and select a random distribution
of coherent state amplitudes consistent with this probability
distribution.  The random realizations thus obtained support
our picture of quantum noise providing the seeds for the
formation of domains, and reproduce the results of Saito et
al.\cite{Saito:2007a}.

The thermal nature  of the fluctuations may be verifiable experimentally. An experiment could be undertaken that considers two settings:\emph{(i)} a system with a starting state of $N$ particles that have Bose-condensed in a ground state of the system, say with $m_{F_z}=1$ (any uniform rotation of such a state would do, since we are selecting one of the degenerate set of ground states for the system). For such an initial state, usually approximated as a coherent state, the number variance (or a corresponding density-density correlator\cite{Glauber:1963,Walls:1995,Altman:2004})
for the particles in the $m_{F_z}=+1$ (or $-1$) state is
Poissonian in nature, i.e. $\sigma_{\mu\bf k=0}\sim{N}$.  The density-density correlator for non-zero wavevector ${\bf q}$, is calculated for particles in the $F_z=1$ state and corresponds to having Poissonian shot noise:
\ba
C_{+\bf q}&\equiv&_{F_z=+1}\bra{N}\rho_{\bf q}\rho_{\bf -q}\ket{N}_{F_z=+1}\nonumber\\
&=&N=\overline{N'}_z,
\ea
where $\rho_{\bf q}\equiv \sum_{\bf k}a^\dagger_{1\bf k-q}a_{1\bf k}$ is the time independent density operator for $m_F=+1$ particles and $\overline{N'}_z$ is the average number of these particles in this starting state.  The correlator is  independent of wavevector $\bf{q}$,  as well as being time-independent - since we begin in a ground state that has no dynamics.

%:figure:correlation
\begin{figure}[t]
\includegraphics[width=3.375in]{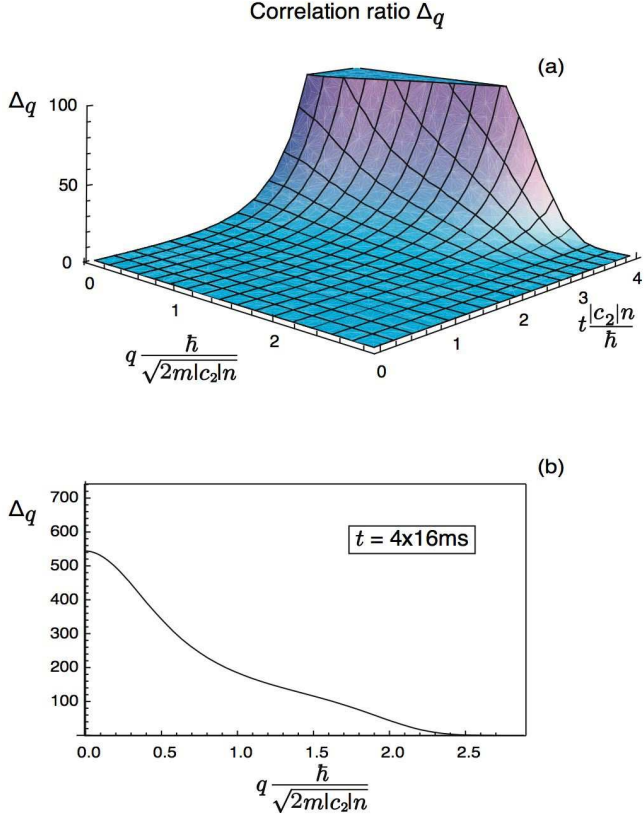}
\caption{\label{correlation} (Color online) The graphs show the correlation $\Delta_{\bf q}$ for: (a) a range of times and wavevectors and (b) for a given time of $t=4\times 16 ms$.  At this time the signal to Poissonian noise ratio is ${\mathcal O}(10^2)$.}
\end{figure}

\emph{(ii)} In the second proposed setting, the starting state would be that corresponding to the Berkeley experiments\cite{Sadler:2006}, and modeled in this paper, with $N$ particles in the $m_{F_z}=0$ state. This we predict would exhibit a thermal distribution for the particles
in the $m_{F}=+1$ state, showing super-Poissonian statistics. The density-density correlator for non-zero wavevector ${\bf q}$ for this state is time dependent, and we use the time dependent creation operators already derived in this paper: 
\ba
C_{0\bf q}&\equiv&_{F_z=+1}\bra{N}\rho_{\bf q}(t)\rho_{\bf-q}(t)\ket{N}_{F_z=+1}\nonumber\\
&=&\sum_{\bf k} \overline{N}_{1\bf k}(1+\overline{N}_{1 \bf k-q}),
\ea
where $\rho_{\bf q}(t)\equiv \sum_{\bf k}a^\dagger_{1\bf k-q}(t)a_{1\bf k}(t)$ is the time dependent density operator for $m_{F_z}=+1$ particles, and $\overline{N}_{1\bf k}$ was defined in Eq.(\ref{Nkavg}). The comparison of the two settings, by looking at the ratio of the experimentally computed correlations for each case and comparing to the theoretical predictions, would provide a direct verification of the unique statistics associated with the two-mode squeezing in these $F=1$ spinor condensates. We may define a correlation ratio deviation, $\Delta_{\bf q}$, implicitly,
\ba
\frac{C_{0\bf q}}{{C_{+\bf q}}}=1+\Delta_{\bf q}.
\ea
The correlation ratio deviation shows us how much the noise differs from that of a coherent state and corresponds to a signal to noise figure for an experiment. For our model we may calculate numerically $\Delta_{\bf q}$ as a function of time, using for $C_{+\bf q}$ the number of $m_F=+1$ particles present in our sample at a given time, having been produced from our initial $m_F=0$ source,
\be
\Delta_{\bf q}=\frac{\sum{\overline{N}_{1\bf k}\overline{N}_{1 \bf k-q}}}{\sum_k N_{1\bf k}}.
\ee

%:figure:correlationRatio
\begin{figure}[tb]
\includegraphics[width=3.375in]{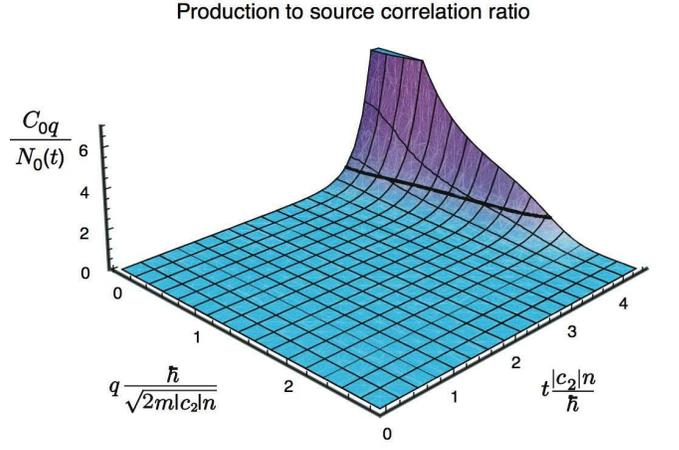}
\caption{\label{correlationRatio} (Color online) The ratio $C_{0\bf q}/N_0(t)$ grows above unity, at later times (indicated by the region above the black curve).}
\end{figure}

Our calculations, shown in Fig.~(\ref{correlation}), indicate an  expected $\Delta_{\bf q}$ to be $\mathcal{O}(10^2)$ for initial times $t\sim 4\times16$ms.  This suggests that the thermal noise should be readily observable in a typical experimental setting.  Furthermore if we look at the ratio $C_{0\bf q}/N_0(t)$, Fig.(\ref{correlationRatio}), which is the ratio of the density-density correlator for our starting state to the Poissonian density-density correlator expected for the number of $m_F=0$ source particles remaining in the sample at any given time, $N_0(t)$, we see that, say at times $\sim 4\times 16$ms, this is greater than unity.  This suggests that at this time, and onwards, the thermal correlations are greater than the Poissonian system shot noise and should be experimentally observable.  Such an experiment to probe the correlations in Bose-Einstein condensed gases has been proposed by Altman et al.\cite{Altman:2004}. The experimental effect of super-Poissonian statistics is reminiscent of the Hanbury-Twiss-Brown effects\cite{Brown:1956}, and the varying number statistics in our case are due to the perfect
correlations between $m_F=\pm1$ pairs in the Berkeley
experiments.

\section{Summary}
In this paper we have discussed the seeding of domains in the
magnetic system of a Bose-Einstein spinor condensate of
$^{87}$Rb atoms, in the spin triplet $F=1$.  The analysis was
performed in the context of the experiment performed by
Sadler et al.\cite{Sadler:2006}.  The spinor gas behaves
ferromagnetically, and the experiment showed the emergence of
domains of transverse magnetization evolving from an initial
polar state in the $m_F=0$, that becomes dynamically unstable
as the initial magnetic field is quickly ramped down.

We have described the dynamical quantum fluctuations of such a
sample that starts as a condensate of $N$ atoms in a pure $F=1$,
$m_F = 0$ in analogy to the `two-mode squeezing' of quantum
optics. Specifically we have modeled the initial $m_F = 0$
condensate as a source for the creation of particle pairs in
the $m_F =\pm 1$ states.  Our considerations for the
single-mode Hamiltonian were extended to a multi-mode
approximation which lead us to a representation of the system
via an $\mathfrak{su}$(1,1) algebra.   Even though the system
as a whole is described by a pure state with zero entropy, it
turned out that  considering only one species at a given
wavevector at a time we obtain super-Poissonian fluctuations
that correspond to a thermal state.  This may be thought of as
equivalent to considering the reduced density matrix for the
$m_F = 1$ degree of freedom, by tracing out the $m_F =\{-1,0\}$
degrees of freedom. These quantum fluctuations of the initial
dynamics of the system provide the seeds for the formation of
domains of ferromagnetically aligned spins. The super-Poissonian fluctuations should be observable experimentally  by looking at density-density correlations to compare our time-evolved states to coherent states.

We should finally note that if we were to use a mean-field
Gross-Pitaevskii approach then we would not observe the
evolution of the system in the absence of noise.   Our approach
provides the source of the instability needed for the
Gross-Pitaevskii equation, in terms of the quantum noise of
squeezed quantum states.  As an improvement  on the standard
mean field description of the system, one may  try to match the
fluctuations of the quantum states to the initial boundary
conditions for the mean-field Gross-Pitaevski equations - such
an approach has appeared in the work of Saito et
al.\cite{Saito:2007a}.  This provides a semi-classical but
fully non-linear description that may probe the system in the
metastable time regime, after the domain formation is complete,
to investigate their stability and evolution. The evolution of
the uniform starting state into domains may be viewed a process
akin to the Kibble-Zurek mechanism\cite{Damski:2007,Kibble:1976,Zurek:1985,Saito:2007a} originally developed to describe the initial evolution of cosmological
defects in the early Universe. Furthermore the processes described have analogues in different models for the production and statistics of particles in cosmology \cite{Guth:1985,Hamilton:2004,Hawking:1971}.  It may be possible in the future to even make the connection more explicit and perhaps
even get to the point of using Bose-Einstein spinor condensates
as analogue models for cosmological
considerations\cite{Volovik:2001,Volovik:2001a,Volovik:2003}.

\begin{acknowledgments}
We are grateful for helpful conversations with M.~Baraban, F.~Song and L.~Glazman. GIM and SMG were supported by NSF DMR-0603369 and NRC was supported by EPSRC grant GR/S61263/01.
\end{acknowledgments}

\end{document}